# A Unified Approach of Detecting Misleading Images via Tracing its Instances on Web and Analysing its Past Context for the Verification of Content

Deepika Varshney, Dinesh Kumar Vishwakarma
Biometric Research Laboratory, Department of Information Technology, Delhi Technological University, Delhi-110042

**Abstract**

The verification of multimedia content over social media is one of the challenging and crucial issues in the current scenario and gaining prominence in an age where user-generated content and online social web platforms are the leading sources in shaping and propagating news stories. As these sources allow users to share their opinions without restriction, opportunistic users often post misleading/ unreliable content on social media such as Twitter, Facebook, etc. At present, to lure users towards the news story, the text is often attached with some multimedia content (images/videos/audios). Verifying these contents to maintain the credibility and reliability of social media information is of paramount importance. Motivated by this, we proposed a generalized system that supports the automatic classification of images into credible or misleading. In this paper, we investigated machine learning-based as well as deep learning-based approaches utilized to verify misleading multimedia content, where the available image traces are used to identify the credibility of the content. The experiment is performed on the real-world dataset (Media-eval-2015 dataset) collected from Twitter. It also demonstrates the efficiency of our proposed approach and features using both Machine and Deep Learning Model (Bi-directional LSTM). The experiment result reveals that the Microsoft bings image search engine is quite effective in retrieving titles and performs better than our study's Google image search engine. It also shows that gathering clues from attached multimedia content (image) is more effective than detecting only posted content-based features.

## 1. Introduction

Nowadays, online social media (Twitter, Facebook, YouTube, etc) is one of the crucial and popular mediums of sharing an individual's thoughts and opinions regarding some event. User can freely share their emotions what he/she think about a certain situation. This open sharing of thoughts and opinions can be a good way of moving information from one to another, but if it can be utilized for malicious purposes (for spreading false information/rumors) to mislead people, it can be a curse for society[1]. In the current pandemic COVID-19, people have their eye on any news article related to covid cure, lockdowns, and other related information. Some people use it as a stepping stone to spread false information for various reasons, either to mislead people, for some monetary benefits, on behalf of some political propaganda, etc. Social media platforms like Twitter and Facebook are prominently used platforms for news diffusion and offer possibilities for rapidly disseminating news to one's zone of contacts and broader communities. This is especially true in those cases when the multimedia content is also associated with the claim. People more prominently share the content rapidly; those with some multimedia item(images/videos/audio) are attached to validate the claim. These posts are often undergoing faster and wider sharing and also going viral. Due to a high volume of content generation and its propagation creates a big challenge for journalists to process the information. There may also be a risk of accepting some false information as true.

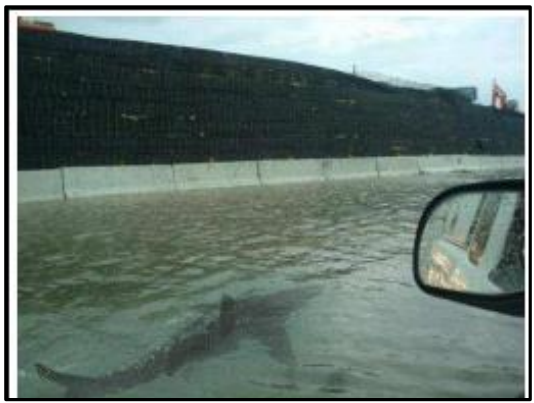
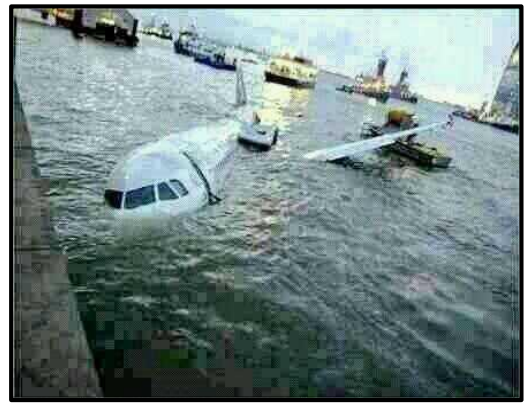

(a)  (b)

Fig.1 Example of Fake news (a) shows the fake photograph that is manipulated and propagated during hurricanes sandy depicting the shark swimming in a flooded freeway and (b) shows the photograph that is wrongly used to represent that the plane is MH370 and raising the false alarm that the plane was detected.

In this paper, we followed the definition of misleading content provided by [2] "*A misleading content can be defined as a content/ claim attached with some multimedia item that does not faithfully represent the event that it refers to*". There can be different cases concerning this (a) some content from the past event is reposted in the context of some similar currently happening event, (b) content that is manipulated/ tampered, and lastly, (c) a multimedia content that posted together with a false claim about the presented event. On the other hand, the post shares the

tweet/ claim that faithfully represents the accompanying multimedia item considered as real posts. Fig.1(a) shows the fake photograph that is manipulated and spread widely during hurricanes sandy depicting the shark swimming in a flooded freeway. In the same way, there is another fake story disseminated. After the disappearance of Malaysia Airlines Flight MH370 in March 2014, lots of false information and fake images are spread on social media and raising a false alarm that the plane was detected[1] as shown in Fig.1(b). This news hampers public emotions that are directly involved in the incident, such as passenger's families. They give an example that concerns the development and need for the technique to identify misleading content. Some earlier approaches utilize content verification by employing Exif metadata of content that incorporates the information related to the date, time, and location where the image was taken [3]. The tweet-based features, user-based features [2], forensics feature [4] predict whether the image accompanying a claim/tweet is credible and faithfully represents an event. All the work, as mentioned earlier are more rely on post-related clues. However, none of the work utilized images for getting efficient clues from the available instance on the web. From the analysis, the existing traces available on the web play a major role in predicting whether the image is pretending the claim/tweet in the correct context and improving the model's performance. Often, the image of some event may be utilized to present in some other context to create chaos and confusion among the public. The main contribution of this paper includes the following:

- We present a new method of predicting misleading content that incorporates image as an accompanying multimedia item and proposes five novel features (Trace of fake concerning to query, Trace of fake concerning to titles, Trace of doubt concerning to query, Trace of doubt on titles, the semantic similarity between title and a query) concerning tweet and images.
- The proposed approach utilizes the images instead of relying on the tweet to retrieve evidence for the prediction of fake, where firstly it includes gathering significant clues from an image via tracing it on an image search engine and then collecting its past instances to retrieve the relevant crucial knowledge for prediction using both deep and machine learning models.

---

[1] snopes.com/photos/airplane/malaysia.asp.

- Prominently used search engines (Google image search and Microsoft Bing visual search) are observed. It has been found that the Bing visual search is quite better for retrieving effective titles and performing better than google images.
- The comparative study is performed on the Mediaeval VMU 2015 dataset, and the proposed method outperforms other state-of-the-art methods.

The rest of the paper is organized as follows: Section 2, describes the related work, whereas Section 3, describes the problem statement. Section 4, explains the proposed methodology for the prediction and classification of a post. Whereas Section 5, gives a detailed experimental analysis and summarizes the results, later we conclude with some future work.

## 2. Related Work

The proposed work mainly focuses on the crucial problem of detecting misleading posts on social media and, more specifically, related to Twitter posts. The tweet/claim accompanied by some multimedia item, either an image or a video, is attached in support to validate the claim. Detecting misleading information is a quite similar concern related to other interesting problems ranging from spam detection [5] to clickbait detection [6], rumor detection [7], satire detection, fake news detection[8] , hoax [9] etc. However, the above problems are distinct in the following ways. For example, Hoax detection is the most commonly used combination of database cross inspecting and reasoning for verifying claim/tweet. In the same way, rumor detection utilizes social media content but employs a collection of posts. In contrast, the main aim of this paper is to verify individual social media posts, typically posted in the context of an unfolding newsworthy event. When a multimedia post is disseminated over social media, very little / no contextual clues are available that can help in predicting the post is misleading or real.

The problem that we are covering in this paper is the focus of VMU the 2015 benchmark challenge. The main aim of the task is to predict the credibility of a multimedia post. The meaning of multimedia post is the post that incorporates tweet and the accompanying multimedia item (image or video) concerning some event, return a binary decision showing the credibility of whether the attached multimedia content faithfully reflects the reality of the situation/ event in the way presented by the tweet. From the previous studies, it has been observed that many authors have to employ text-based features from the post, and classification has been done using machine learning algorithm, text-based like the presence of punctuation, language style, linguistic patterns. Another feature that has been explored is user-based i.e., knowledge is extracted from user profile/ account who made the post like the number of followers/friends, or interaction-based, age or number of followers/friends.

Many of the previous work reported techniques to detect variants of fake [clickbait's, rumors, fake news, hoax, etc.][10]. One of the first and earliest studies on assessing content credibility at the event/topic level is provided by [11]. The author employed text-based, user-based, and topic-based features. In the same way, some of the studies worked on linguistic-based features retrieved from news stories' textual information. The authors proposed a set of linguistic features like positive/negative sentiment of words, emojis, etc. to predict fake news. Whereas, in [12], the author utilizes language stylistic features like assertive verbs, discourse markers, etc. to assess the credibility of a post. Similarly, the authors of [13], employ text, user, and propagation-based features. Some of the authors employ deep neural networks and explore the possibility of showing tweets concerning the deep neural network. In [14], attention mechanisms have been incorporated into a recurrent neural network (RNNs) to extract distinct temporal linguistic features with a particular focus. The authors of [15], proposed tweet level features to classify tweet sharing fake images and tweet sharing real images on one of the datasets of tweets related to a Hurricane Sandy event. In this way, classification is used to verify the accompanying images, and this study is quite related to our work. Similarly, the author of [2], proposed an effective framework for predicting Twitter post whether it is fake or real by employing a publicly available verification corpus. The proposed features based on tweets and user are effective in improving the performance of the model. The use of bagging and the application of an agreement based retraining approach are effective and outperforms standard supervised learning. Whereas, the author of [16], proposed a framework (Multimodal variational autoencoder for the task of detecting fake news, where the model incorporates three major components, an encoder, fake news detector module and a decoder. Similarly, from one study it has been reported that the image associated with some claim, play a crucial role in differentiating fake from real posts, as it has been seen that they have distinct visual characteristics [17],[18]. To get traces of fake from attached multimedia item [images/ videos], the authors are also keen their interest towards multimedia forensics, to identify any traces of manipulation/tampering in the image [19],[20],[21] and videos [22]. There are prominently used techniques such as splicing detection [21], copy-move forgery detection [20]. However, these methods are not well suited for social media images as it is very likely that the image conveys false information without being manipulated/ forged. For example, the image of some authentic past event may be used to misrepresent some current event in context. So the major aim is to get the effective traces from both tweets and from the associated image that can faithfully validate the post. The novelty of our proposed work is shown in Fig.2. Fig.2 shows that the clues are extracted from the multimedia posts from two parts, including image and text. The effective clues are fetched by incorporating Tweet and Image together and via getting clues from images themselves. In most previous studies, researchers utilized image forensic features like image color, checking for any tampering/ manipulations, etc. But sometimes, an image may be not tampered with/ manipulated, however, the image is wrongly attached with a claim/item representing something in a different context to mislead people, and in that case, applying any forensic technique is not applicable. To address these cases, we have incorporated a novel mechanism to gather clues by tracing an image on the web and identifying its past context, analyzing the content that is useful in fetching efficient clues.

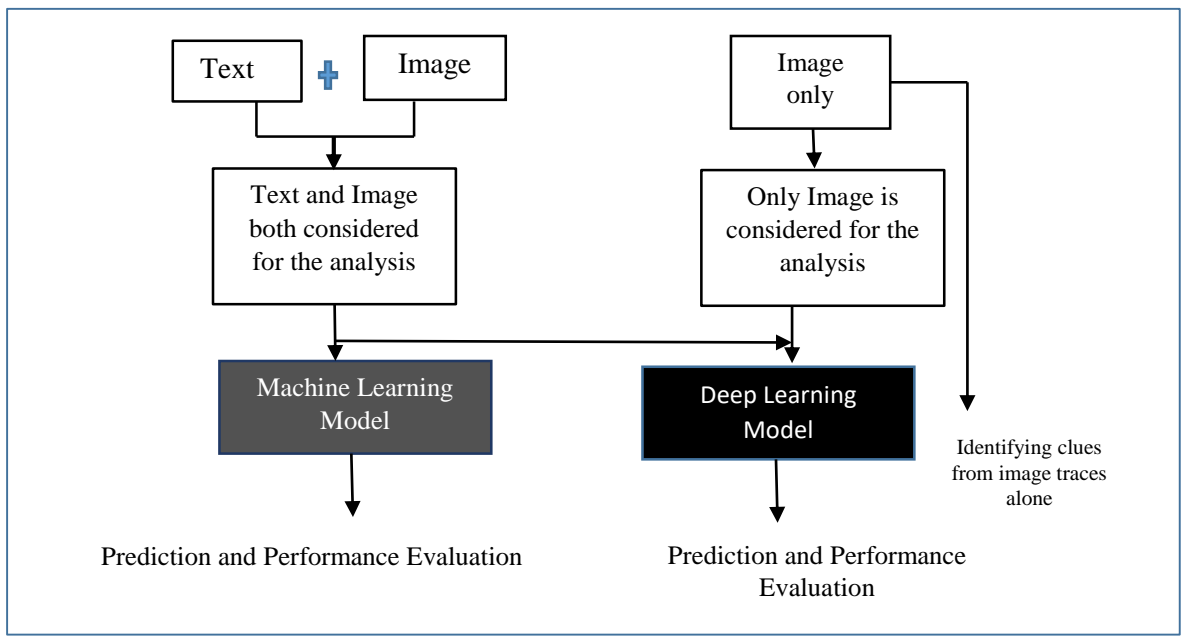

Fig.2 The novelty of the proposed work

## 3. Problem Description

In this section, we describe the problem description and briefly explain the generalized model for the verification of multimedia content posted on social media. The multimedia post we have considered here is incorporating two parts 1. Image Part 2. Tweet/claim Part. In this work, any post associated with these two parts is considered as a *multimedia post*, the detailed description is given in the following section.

### a) General Overview

In this paper, the efficient clues that have been retrieved for the prediction of misleading content are from two parts of any multimedia post: Tweet + Image, and Image only. The first part incorporates both tweets and images for the retrieval of efficient clues. The important evidence considered in this category is based on semantic similarity, fake and doubt traces that further be used for the classification using machine learning models. The feature-based evidence concerning each multimedia post can be represented as $m^i = (DB^i, UNS^i, S^i)$ Where, $DB^i$ defines that the user is in doubt with the claim accompanying multimedia item(image) from an event. The $UNS^i$ defines that the user does not support the claim and is not confident with the accompanying multimedia item(image) from an event. Whereas, $S^i$ defines the semantic similarity score. These crucial factors are identified concerning each multimedia post for the prediction of misinformation. Including this set of clues/factors, the other deep learning aspect has also been explored concerning the Tweet + Image part. The hidden representation of word sequences has been generated using the Bi-directional LSTM model for the prediction of misleading content. The concept is discussed in detail in the later sections.

The second part (Image only) of analysis has been applied by extracting crucial knowledge from the existing instances of an image found on the web. Here, we have only considered the image traces retrieved in the form of title/ headlines (top 10) concerning each image using the google reverse image technique that further goes as an input to the Bi-LSTM model to get the hidden representation from the text. This case is effective when we have only an image as an input and we need to predict whether an image is misleading or not. It has been observed from the empirical analysis that for some of the images, relevant claims are not retrieved or the google search engine is not able to identify the images in the correct context. Due to this, useful search results may not be retrieved. To resolve such a scenario, we have also retrieved traces of an image from another prominently used search engine i.e Microsoft bing visual search[2]. Some of the results responses from Microsoft visual search and Google image search[3] concerning an image have been shown in Table 2. The results reveal that Bings visual search gives quite better and relevant responses in context to an event compared to google image search responses in our study. We will discuss the detailed comparative study of both Image search engines in the later section.

---

[2] See it, search it | Bing Visual Search
[3] https://images.google.com/

## b) Aim/ Objective

Each multimedia post is associated with '$n$' claims posted by '$m$' users. Multiple users share their different opinion concerning an individual image. The aim is to verify the given claim/tweet and the accompanying multimedia item(image) from an event that they are faithfully describing each other and not contradictory, further return a binary decision representing verification of whether the multimedia item reflects the reality of the event in the way purported by the tweet.

In this study, we have considered 'n' events and there are 'm' multimedia posts concerning each event. For each multimedia post, there are 'r' users showing their point of expression/ opinion by posing 'k' claims. We can show the complete scenario and relationship between different object modules of our system.

The detailed description of each of these two-part has been discussed in Section 4.

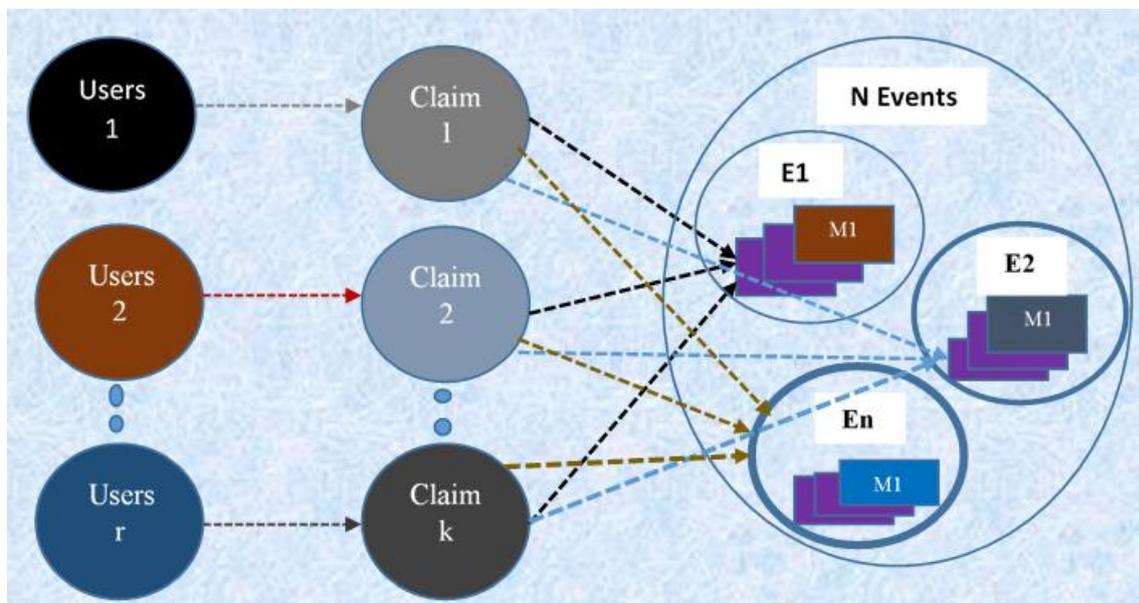

Fig 3. The figure represents the relationship between user, claims, events and an Image.

The graphical representation of our system which is a group of users, claims, events, and an image is shown in Fig.3. The graph clearly shows the relationship among them, where there are a set of *"r"* users posting different opinions about a specific multimedia post related to some event. There are "*N*" events, each event accompanying a *"k"* multimedia post. Opinions give a set of claims that a user is thinking about the specific event and expressing their thoughts to represent the given situation. Most of the time, on social media, people share thoughts without verification, just that post goes viral, people are supporting the given news. While posting any multimedia post, there can be multiple possible cases that can be applied with respect to the human point of expression.

- The user is in support of the claim and confident with the accompanying multimedia item(image) from an event. This we termed as confident claims $CON^{(i)}$.
- The user is in doubt with the claim and with the accompanying multimedia item(image) from an event. This we termed as doubtful claims $DB^{(i)}$.
- User is not in support of the claim and not confident with the accompanying multimedia item(image) from an event, which is termed as unsupportive claims $UNS^{(i)}$

By understanding the human point of expression, we can evaluate the uncertainty score of the claims provided by users on a specific event and can observe user's expressions using Eq.1. We can evaluate the uncertainty score. The uncertainty score can be calculated as the Boolean sum of DB and UNS value for the i$^{th}$ tweet/claim. There is a list of phrases and a corpus of words is created from the empirical analysis of collected data. For the doubtful claims, we are analyzing whether the tweet contains any question marks. Question marks are an effective way of identifying the user expression that he/she is in doubt with the given accompanying multimedia content and it represents the uncertainty in their opinion. If any question mark has been identified in the tweet, the DB value will be 1 and 0 otherwise.

$$uncertainty\_score(CS) = (DB^{(i)} + UNS^{(i)}) \qquad (1)$$

## 4. Evidential Clues for the Verification of Misleading Multimedia Content

Selecting and incorporating the right set of features and input parameters plays an important role in the better performance of the model. The effective features have been extracted from the multimedia post that leads to give efficient clues for the prediction of misleading content.

4.1 Evidence Collection from (Tweet Part+ Image) part using Machine learning models:

In this section, we are going to cover the set of evidence or clues that have been collected from the tweet as well as from an image. In this study, we have considered multimedia posts with a claim/tweet and the accompanying multimedia item(image). The available tweets are in multilingual form, to understand the semantics, language translation has been applied using google trans library of python. Google trans is a free and unlimited python library that implemented Google Translate API[4]. After analyzing the tweet, it has been observed that the pattern of question marks and trace of false phrases can be an efficient clue for the prediction of false information. Before going to discuss the clues related to an image, let's first discuss how we can process an image to retrieve relevant knowledge?. In our proposed idea, any multimedia post attached with an image is processed as follows, the associated image is given as an input to the image search engines (i.e., Google Image search and Bings visual search here) and each search engine returned relevant available instances/ context matching with an image. So, in this case, the verification of result responses, whether they are related to the search query is not necessary, because here by default we are getting only those instances on the web, having correlated images. The retrieved titles from each image were further used to gather clues. The following measures considered both tweets and images to gather efficient clues for the prediction of misleading information.

4.11 Trace of Doubt(DB):

This is one of the patterns that widely identified in the human expressing pattern when he/she is in doubt regarding what they are posting and not sure regarding the post. After analyzing the dataset, we built a corpus having phrases concerning to trace of doubt. We observed that the prominently used words for expressing doubts are {*is it, is that, Not sure, ?*}. The return value is binary, if it returns 1 means that the tweet expressing doubt, otherwise 0. Here we have represented the trace of doubt with the term DB as discussed in Section 3.

4.12 Trace of Fake(UNS):

Trace of fake is another pattern we have analyzed in the tweets, where the user itself showing the expression of fake and presenting that they are not supporting the claim. We have built a corpus*{'Malware', 'Beware', 'scam', 'fishy', 'phishing', 'funny', 'Not', 'ambiguous',' false', 'misleading', 'inaccurate', 'rumor', 'rumour', 'fool', 'fooled', 'not correct', 'wrongly', 'wrong',' misidentified', 'fake news', 'falsely', 'incorrect', 'memes', 'catchy', 'bogus', 'fabricated', 'forged', 'fraudulent', 'artificial', 'erroneous', 'faulty', 'improper', 'invalid', 'invalid', 'mistaken', 'unreal', 'untruthful', 'fishy', 'illusive', 'imaginary', 'lying', 'misrepresentative', 'falsity', 'falsification', 'fabrication', 'falsehood', 'hoax', 'incorrect', 'not real', 'not true', 'fishy', 'illusive', 'imaginary', 'lying', 'misrepresentative', 'falsity', 'misreport', 'deception', 'falsification', 'lie', 'scandal', 'misinformation', 'misleading', 'not* dead', 'death rumor', 'not known', 'no proof', 'no scientific evidence', 'denied', 'deny', 'unverified', 'myth'} of prominently used words pattern in the tweets for representing the trace of fake. Here we have represented the trace of fake with the term UNS as discussed in Section 3. If any of the word patterns have been detected in the tweet it will return 1 otherwise 0.

---
[4] https://pypi.org/project/googletrans/

4.13 Semantic Similarity Measure

The semantic similarity between a tweet and the titles retrieved from an image search responses ranges from 1 to 10(Top 10 titles) has been calculated. The semantic-text-similarity library of python is an easy-to-use interface to fine-tuned BERT models for computing semantic similarity[5]. This semantic similarity can be one of the good measures to compute how similar the two sentences are contextual. This will also reveal that the posted claim/tweet faithfully represents the accompanying image or not. The Semantic Bert similarity maps batches of sentence pairs to the real-valued scores in the range [0,5]. From the empirical analysis of the similarity value in the dataset, we decide the threshold values that reflect whether the tweet and title are represented in the same context or contradictory or not matched. Table 1, shows the set of possible cases that can be applicable and by empirical analysis on Bert-semantic similarity score we have decided the threshold value T, if $T < 1.3$ it has been observed that the given tweet/claim and title point of expression are not in the same context and contradictory or not matched to each other, for example, suppose the query is "This image is NOT MH370, this is an image from the incident of a plane crashed in Sicily on 6Ogos2005 #PrayForMH370" and the retrieved title is "Atr72 air disaster, Bari remembers 16 victims". The computed semantic similarity value is 1.03 which is less than the threshold value T, and it represents that the title and the query are represented in a different context, whereas, if the T>= 1.3 it shows that the query and tweet are represented in the same context, for example, the query is "*This image is NOT MH370, this is an image from the incident of a plane crashed in Sicily on 6Ogos2005 #PrayForMH371*" and the title is "*Serious! - Pictures of MH370 Crashed at Sea This Is Fake UPDATES*" have T value 2.125, which is more than 1.3. In addition to this, the trace of fake has also been check concerning each query and title that whether they are reporting some expression of fake. Three cases can be possible here as shown in Table1.

---
[5] [semantic-text-similarity · PyPI](#)

The first case is when the Query/ Root itself reporting news as fake, while clue is not reporting the trace of fake and in contradiction or not matched, while the second case says that the Query/ Root is not reporting the trace of fake, while clues are reporting and in contradiction and the third case is Query/ Root and clue both are reporting the news as fake and in support of each other. In Fig.4, the Process describing how semantic similarity value between query and clue can be an effective factor classifying fake and real. These set of features are passes to the machine learning model for the prediction of misleading posts as shown in Fig.4.

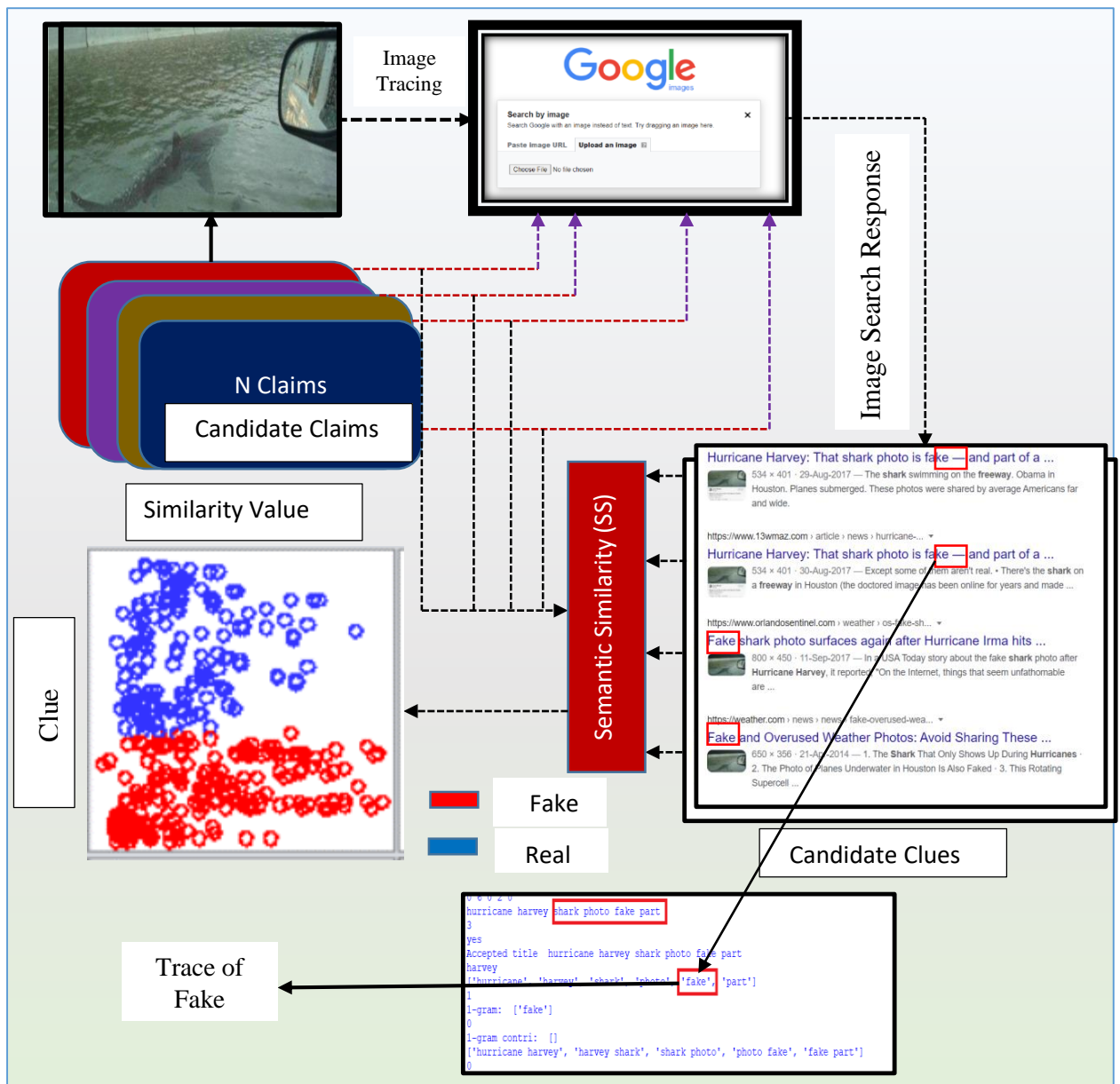

Fig 4. Process describing how semantic similarity value between query and clue can be an effective factor classifying fake and real.

Table1: The set of possible fake cases that can be applicable

| Bert Semantic Similarity value(T) | Identified Fake Cases | Query False Phrases | Clue False Phrases |
|---|---|---|---|
| T<1.3 | Context is not the same | - | - |
| T>=1.3 | Query/ Root itself reporting news as fake, while clue is not reporting the trace of fake and in contradiction | Yes | No |
| T>=1.3 | Query/ Root is not reporting the trace of fake, while clues are reporting and in contradiction. | No | Yes |
| T>=1.3 | Query/ Root and clue both are reporting the news as fake and in support of each other. | Yes | Yes |

Table 2. Image search result responses from Microsoft bings and google image search.

| Images | Microsoft Bings Image Search Responses | Google Image Search Responses |
|---|---|---|
| 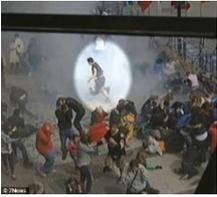 | <ul><li>Now: FBI Hunts Suspect in Boston \| Bomb Attack.</li><li>Common Cents: Are these photos of the Boston Bombing Suspect?</li><li>Photo of Boston Bomber Caught on camera \| TODAY'S JOBS \|.</li><li>Boston Marathon suspects Archives</li><li>Who of these men is the Boston Marathon Attacker? Possible Suspect in.</li></ul> | <ul><li>event web apis mdn</li><li>event meaning Cambridge English dictionary</li><li>search results signals az.</li><li>digital vigilantism boson marathon bombing</li><li>boston marathon bombings latest arrest made.</li></ul> |
| 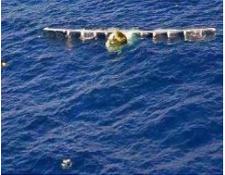 | <ul><li>Atr72 air disaster, Bari remembers the 16 victims</li><li>Cape Gallo air disaster, 11 years ago the Atr72 tragedy.</li></ul> | <ul><li>Crash pilot who paused to pray is convicted \| Reuters</li></ul> |
| 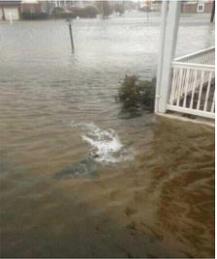 | <ul><li>Is that picture real or **fake?** - Is that right?</li><li>20 Epic **Fake** Pictures that Have **Fooled** the Whole World \| Shark swimming.</li><li>The Big Apple has lots of sharks. But real ones in the neighbourhood.</li><li>Super Storm Sandy Sharks swimming down New Jersey street.</li><li>Hurricane Irene: 'Photo' of shark swimming in street is **fake** \| Shark .</li></ul> | <ul><li>54 Super storm sandy ideas \| sandy, storm, hurricane sandy.</li><li>72 Crazy shit ideas \| hurricane sandy, natural disasters, photo.</li><li>7 Sandy ideas \| sandy, hurricane sandy, hurricane pictures.</li><li>These Viral Shark Photos from Hurricane Matthew Are, Once.</li><li>**Fake** and Overused Weather Photos: Avoid Sharing These.</li></ul> |
| 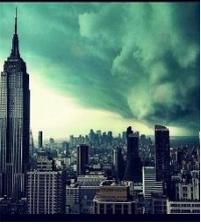 | <ul><li>**Is that** really a picture of Hurricane Sandy descending on New York**.?**</li><li>NY City \| Hurricane pictures, New York photos, New york \|</li><li>Hurricane Sandy 2012: 10 Amazing Photos of the Storm's Path Through New.</li><li>These Are **NOT** Photos From Hurricane Sandy (No Matter What The Internet.</li></ul> | <ul><li>Internet Awash in #**Fake** Sandy Photos. Have You Shared Any?</li><li>22 Viral Pictures That Were Actually **Fake** \| Hurricane pictures ...</li><li>Example of **fake** picture of stormy New York skyline used in ...</li></ul> |

4.2 Detect complex patterns from (Tweet and Image) and (Image only) search responses using Bi-directional LSTM

To Extract the complex hidden representation from tweets and Image search responses, a Bi-directional Long short-term memory network (Bi-LSTM) a special type of RNN competent in learning long dependencies is utilized in our proposed work as shown in Fig.5. An RNN has an internal state whose output at every time step can be expressed in terms of the previous time step. However, RNNs suffer from the problem of vanishing and exploding gradients[6] and this leads to the model learning inefficient dependencies between words that are a few steps apart. To overcome this issue, the LSTM extends the basic RNN by storing information over long periods by its use of memory units and efficient gating mechanisms. LSTM is a special type of RNN competent in learning long-term dependencies and they are providing an efficient solution to address the vanishing gradient problem. In LSTM-RNN the hidden layer of basic RNN is replaced by an LSTM cell. LSTM is prominent as they utilize various gates in their architecture that help in learning how and when to forget and when not to. Another variant of RNN is Bi-directional LSTM, where you feed the learning algorithm with the given data in two ways once from beginning to the end and once from end to the beginning. From the study, it has been observed that for a large text sequence prediction and text classification, Bi-directional LSTM was found to be an effective and evident approach, which takes a step through the input sequence in both directions at the same time. The proposed misleading content detection model is based on Bi-directional LSTM – recurrent neural network. The tweets/claim and the image search responses(Titles) corresponding to each image are first pre-processed (removing stop-words, stemming, lemmatization, removing URLs, punctuation). Concerning each image, there are n responses retrieved (n titles). A binary label is set to each title as 1 for fake news and 0 for real news corresponding to the individual query. The titles retrieved from image search responses and the corresponding query are turned into a space-separated padded sequence of words. These sequences are further split into tokens. One hot vector encoding embeddings is utilized to represent each word by the real value number. The embeddings are then passed to Bi-directional LSTM Model to detect complex hidden patterns/features from the text. The transformed vector represented data is partitioned into train, validation, and test data. The training is carried out on the build corpus of queries and titles concatenated with a space. Validation data set is used for fine-tuning the model. Further, the test data is used to know the predicated label of news content (query + title) based on the trained model. In the same way, the analysis has been applied by considered the traces from an image only and no tweet content has been included. The past context fetched from the web searches concerning an image is utilized to extract the hidden representation from content retrieved through returned responses(title) corresponding to an image. Further, the test data is used to know the predicted label of news content(title) based on a trained model. To minimize the loss function, the model is trained iteratively to improve accuracy. The binary cross-entropy loss is considered to detect misleading multimedia posts in the proposed model. The Adam optimization algorithm is used to improve the performance of the model.

---

[6] Recurrent Neural Networks (RNN) - The Vanishing Gradient Problem - Blogs SuperDataScience - Machine Learning | AI | Data Science Career | Analytics | Success

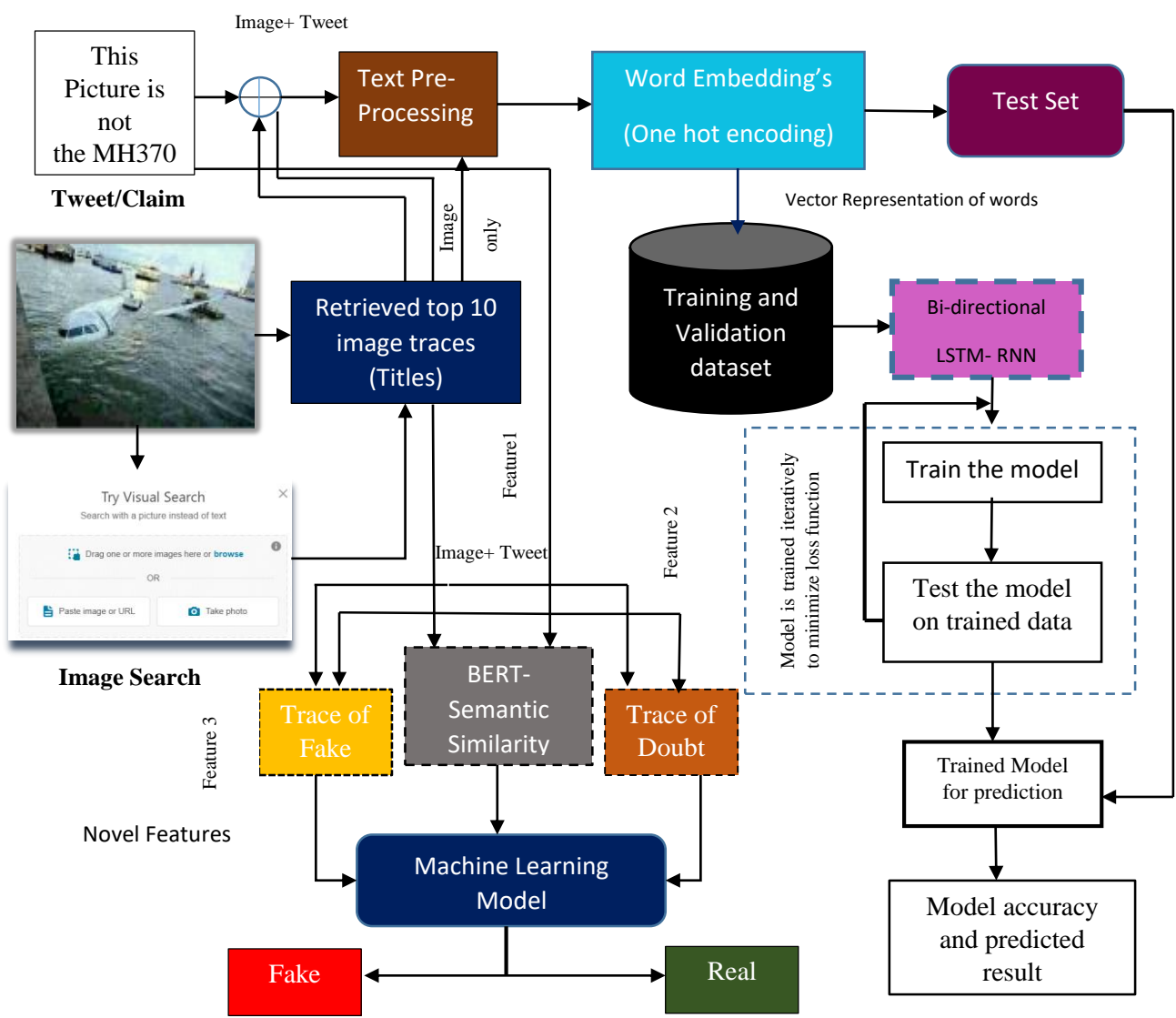

Fig. 5 The Proposed architecture to Detect complex patterns from tweet and Image search responses using Bi-directional LSTM (Deep learning) and machine learning models.

# 5. Experiments and Results

This section discusses the experimental analysis and later demonstrates the results we have achieved by applying our proposed approach for the detection of misleading content on social media. We then briefly discuss some state-of-the-art techniques used in this field and lastly show the comparative analysis with baselines to validate the performance of our model.

## 5.1 Dataset

In this section, we discuss the dataset that has been employed to evaluate the performance of the model. One of the prominently used standard datasets is the Mediaeval verifying multimedia Use challenge[7]. The task was aimed to predict the misleading multimedia content on social media. The dataset is comprised of a set of Twitter posts having tweets associated with multimedia items. The VMU(Verifying Multimedia Use 2015) is a publicly available dataset [23] on GitHub[8]. The dataset incorporated social media posts having ~400 images (176 cases of real and 185 cases of misleading images) associated with 5,008 real and 7,032 fake tweets concerning 11 events (Boston Marathon bombing, Hurricane Sandy, etc.). Table 3 shows the detailed description of the VMU 2015 dataset. As in this study our main focus on the images and textual information, that's why the tweets that are associated with videos are filtered out.

Table 3. The table represent the detailed description of VMU 2015 Dataset.

| Event Name | Real Images | Real Tweets | Fake Images | Fake Tweets |
|---|---|---|---|---|
| Hurricane Sandy | 148 | 4,664 | 62 | 5,559 |
| Boston Marathon Bombing | 28 | 344 | 35 | 189 |
| Sochi Olympics | - | - | 26 | 274 |
| MA flight 370 | - | - | 29 | 501 |
| Bring Back Our Girls | - | - | 7 | 131 |
| Columbian Chemicals | - | - | 15 | 185 |
| Passport Hoax | - | - | 2 | 44 |
| Rock Elephant | - | - | 1 | 13 |
| Underwater Bedroom | - | - | 3 | 113 |
| Livr mobile app | - | - | 4 | 9 |
| Pig fish | - | - | 1 | 14 |
| Total | 176 | 5,008 | 185 | 7,032 |

---

[7] Verification (New!) (multimediaeval.org)
[8] https://github.com/MKLab-ITI/image-verification-corpus.

The study is conducted on the machine as well as deep learning approaches by utilizing images, and the combination of Tweet and images. In the following subsection, we separately discuss the effectiveness of employing image only, and both (image and tweet) the ways as well as analyze the performances with respect to each case.

5.2 Performance Evaluation on Machine learning models

The effectiveness of the proposed method has been evaluated by assessing the novel features employed for the prediction of misleading content. The five-set of novel features as discussed in Section 4 (Trace of fake concerning to query, Trace of fake concerning to titles, Trace of doubt concerning to query, Trace of doubt concerning to titles, the semantic similarity between title and a query) with respect to tweet and images are fed into machine learning model to validate how significant these features in improving the performance of the model. The titles concerning an image are retrieved using Microsoft bing visual search and the performance of the model has been evaluated using TP rate, FP rate, Precision, Recall, F1 score, and accuracy as shown in Table 4. From Table 4, it can be observed that Random forest and Linear SVM performing better and outperform all other classifiers with an F1 score of 0.978.

Table 4. Effectiveness of the proposed model using machine learning methods

| Classifier | Performance Measures | | | | | |
|---|---|---|---|---|---|---|
| | TP Rate | FP Rate | Precision | Recall | F- Measure | Accuracy |
| Random Forest | 0.978 | 0.019 | 0.979 | 0.978 | **0.978** | 97.81 |
| Logistic Regression | 0.970 | 0.026 | 0.971 | 0.970 | 0.970 | 96.99 |
| Naïve Bayes | 0.929 | 0.062 | 0.936 | 0.929 | 0.929 | 92.89 |
| Linear SVM | 0.978 | 0.026 | 0.979 | 0.978 | **0.978** | 97.81 |
| K-Nearest Neighbour | 0.967 | 0.029 | 0.968 | 0.967 | 0.967 | 96.72 |

5.3 Performance Evaluation on Deep learning models

To extract complex hidden representation/ features from textual data, the Bi-directional LSTM model has been employed as discussed in Section 4.22. Here, the performance of the proposed model has been evaluated by employing two prominently used image search engines "Google search and Microsoft Bing visual search" for the retrieval of image search responses. It has been observed that getting effective search responses concerning an image is one of the crucial measures in improving the performance of the model. The performance of the model degrades if significant responses/titles have not been retrieved. To validate this point, the comparative study has been performed by employing two prominent image search engines for the retrieval of image search responses on the Mediaeval dataset, the provided study is when only image search responses(titles) are passes to the Bi-directional LSTM model. One of the examples is represented concerning an event *"Boston Marathon Bombing"* as shown in Fig.6.

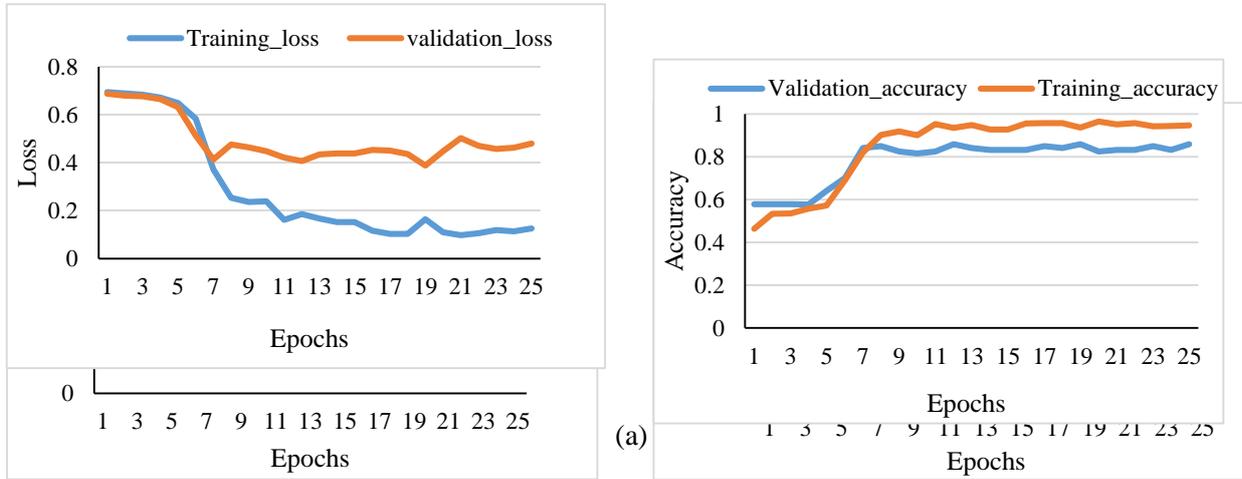

Fig. 6 The training and validation loss as well as accuracy curve corresponding to no. of Epochs for Boston Marathon Bombing. (a) Google Chrome (b) Microsoft Bings (Image only)

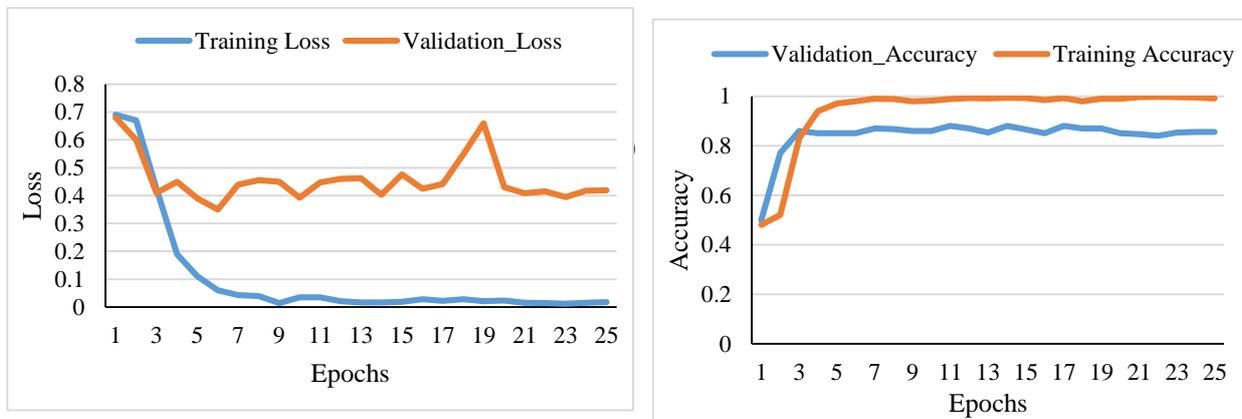

Fig 7. The training and validation loss as well as accuracy curve corresponding to no. of Epochs for overall dataset (VMU 2015) using Microsoft Bings (Image only)

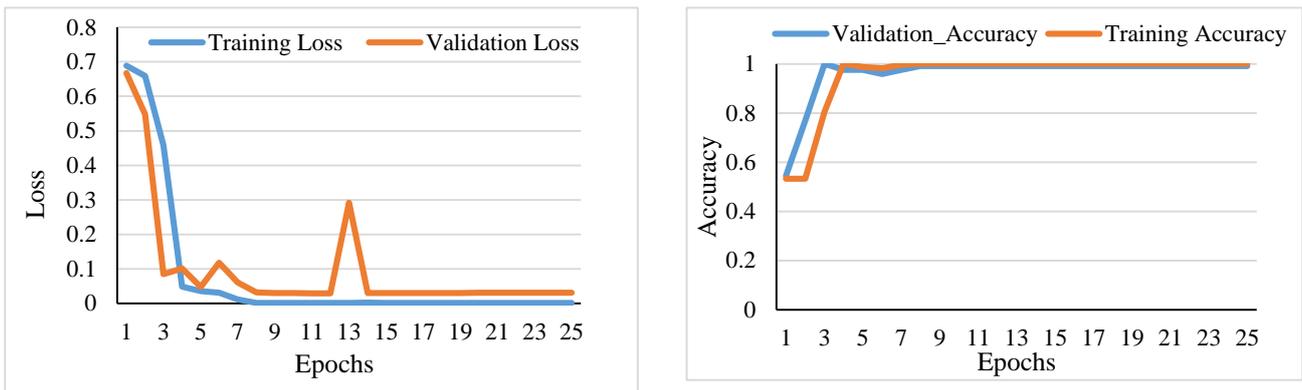

Fig 8. The training and validation loss as well as accuracy curve corresponding to no. of Epochs for overall dataset (VMU 2015) using Microsoft Bings (Image + Tweet only)

The loss and accuracy curve corresponding to the number of epochs is shown to demonstrate the performance of the model. It has been observed from Fig.6(a), that we are achieving the validation accuracy of 0.93 when utilizing Microsoft bings as an image search engine which is quite good and better in comparison when utilizing google chrome image search results (validation accuracy of 0.85) on *"Boston Marathon Bombing"* when reaching 25$^{th}$ epoch as shown in Fig.6(b). From the complete observation, we found that utilizing Microsoft Bing image search is better to improve the performance of our model on our data, that's why incorporated the same for the further analysis. The other set of experiments has been performed on the overall dataset, the provided study is when only image search responses(titles) are passes to the embedding layer and then further passes to the Bi-directional LSTM model. From, Fig.7, it can be seen that we can achieve a validation accuracy of 0.86, and loss is reduced to 0.41. To improve the performance of the model, instead of just passing Image-based clues, the tweet/claim is also incorporated to get effective features. The Tweet and Images search responses are concatenated separately with space and pass to the model. It has been observed from Fig.8. that there is a significant improvement we achieved in this case, we got a validation accuracy of 0.99 and loss is almost reaches 0.

5.4 Comparative study with state-of-the-art approaches

The comparative study has been performed with the other state-of-the-art methods to evaluate the performance of our proposed approach. We compare the techniques applied on Mediaeval VMU dataset 2015 as discussed in section 5.1. From Table 5, it can be observed that the proposed method outperforms the state-of-the-art technique on the same dataset. The main performance measure that has been used for the comparison is F1-score and approaches are compare against their best run. Among all other methods (these include the method proposed by [24], [25], [26], and [27], our method outperforms with an F1 score of 0.99 using the Bi-directional LSTM model and give the best run when considering both tweet and image.

Table 5. The Comparative study between the proposed method and the state-of-the-art method on the medieval VMU 2015 dataset.

| Ref | Method | Type of Input | Performance Measure | | | |
|---|---|---|---|---|---|---|
| | | | Precision | Recall | F1-Score | Accuracy |
| [2] | Logistic Regression | Tweet+ Image | - | - | 0.932 | - |
| [2] | Random Forest | Tweet+ Image | - | - | 0.935 | - |
| [24] | UoS-ITI | Tweet+ Image | - | - | 0.830 | - |
| [25] | MCG-ICT | Tweet+ Image | - | - | 0.942 | - |
| [26] | CERTH-UNITN | Tweet+ Image | - | - | 0.911 | - |
| **Our Method** | LSTM | Image only | 0.86 | 0.86 | 0.86 | 0.86 |
| | LSTM | Tweet+ Image | 0.99 | 0.99 | **0.99** | 0.99 |
| | Random Forest | Tweet + Image | 0.979 | 0.978 | **0.978** | 97.81 |
| | Logistic Regression | Tweet+ Image | 0.971 | 0.970 | **0.970** | 96.99 |
| | Naive Bayes | Tweet+ Image | 0.936 | 0.929 | 0.929 | 92.89 |
| | Linear SVM | Tweet+ Image | 0.979 | 0.978 | **0.978** | 97.81 |
| | K-Nearest Neighbour | Tweet + Image | 0.968 | 0.967 | 0.967 | 96.72 |

However, it gives an F1-Score of 0.86 when utilizing only image-based evidence. The authors of [2], employed supervised machine learning methods for evaluating the performance of their model, where they achieved an F1- score of 0.932 and 0.935 with Logistic Regression and Random Forest respectively. Whereas, by employing our proposed novel features, we achieved an F1-Score of 0.978 and 0.970 with random forest and logistic regression respectively.

## 6. Conclusion and Future Work

In this paper, we have presented a novel and effective method of predicting tweet/ claim accompanying an image to identify how faithfully an image representing a tweet/ claim and to classify them into misleading and real. Using publicly available benchmark verification corpus VMU (2015), we have provided a novel technique via extracting clues from both tweet and image. The five sets of novel clues (Trace of fake concerning to query, Trace of fake concerning to titles, Trace of doubt concerning to query, Trace of doubt concerning to titles, the semantic similarity between title and a query) concerning tweet and images have been extracted from a tweet and images. The images are processed and effective titles are retrieved. From the study, it has been observed that the retrieval of effective titles plays a major role in improving the performance of the model. The two prominent image search engines are utilized for processing an image (Google Image search and Microsoft Bings visual search). From the comparative analysis, it has been observed that utilizing Microsoft Bings Visual Search is quite more effective in retrieving efficient titles and helps in improving the performance of the model. The results showed that the proposed method outperforms the other state-of-the-art methods. In the future, we are more likely to build a solution that can incorporate other multimedia items (Videos, audio, speech attached with tweet/claim) as well as try to build effective real-time application and browser plug-in from a user perspective that can help in the prediction of misleading content in real-time.